# Quantum Oscillations of Robust Topological Surface States up to 50 K in Thick Bulk-insulating Topological Insulator

Weiyao Zhao[a b], Lei Chen[a], Zengji Yue[a b], Zhi Li[b], David Cortie[a b], Michael Fuhrer[b,c],

and Xiaolin Wang[a b*]

[a] *Institute for Superconducting and Electronic Materials, Australian Institute for Innovative Materials, University of Wollongong, NSW 2500, Australia*

[b] *ARC Centre of Excellence in Future Low-Energy Electronics Technologies FLEET, University of Wollongong, NSW 2500, Australia*

[c] *School of Physics & Astronomy, Monash University, 3800 VIC Australia*

As personal electronic devices increasingly rely on cloud computing for energy-intensive calculations, the power consumption associated with the information revolution is rapidly becoming an important environmental issue. Several approaches have been proposed to construct electronic devices with low energy consumption. Among these, the low-dissipation surface states of topological insulators (TIs) are widely employed. To develop TI-based devices, a key factor is the maximum temperature at which the Dirac surface states dominate the transport behavior. Here, we employ Shubnikov-de Haas oscillations (SdH) as a means to study the surface state survival temperature in a high quality vanadium doped $Bi_{1.08}Sn_{0.02}Sb_{0.9}Te_2S$ single crystal system. The temperature and angle dependence of the SdH show that: 1) crystals with different vanadium (V) doping levels are insulating in the 3 – 300 K region; 2) the SdH oscillations show two-dimensional behavior, indicating that the oscillations arise from the pure surface states; and 3) at 50 K, the $V_{0.04}$ single crystals ($V_x$:$Bi_{1.08-x}Sn_{0.02}Sb_{0.9}Te_2S$, where $x = 0.04$) still show clear sign of SdH oscillations, which demonstrate that the surface dominant transport behavior can survive above 50 K. The robust surface states in our V doped single crystal systems provide an ideal platform to study the Dirac fermions and their interaction with other materials above 50 K.

*Corresponding email: xiaolin@uow.edu.au

# 1. Introduction

Since the concept of *topology* has been introduced into condensed matter physics, an exotic form of quantum matter, namely, the topological insulator (TI), has attracted much attention, in which there is no transport of electrons in the bulk of the material, while the edges/surface can support metallic electronic states protected by time-reversal symmetry.[1] Theoretical predictions have revealed spin-moment locking, and linear-dispersion edge states can be addressed in a two-dimensional (2D) topological insulator.[2,3] The experimental breakthrough leading to the 2D system was achieved in a strong spin-orbital coupled CdTe/HgTe quantum well, which gave rise to a quantum spin Hall state.[4] Recently, 1T′-WTe$_2$ monolayer[5-7] has been proved to be a new quantum spin Hall insulator, with a conducting edge state that can survive to 100 K[8,9], thereby offering a potential application for a 2D TI in ultra-low-energy electronics.

In three-dimensional (3D) topological insulators, the conducting edge changes to a metallic surface state that emerges due to the nontrivial $Z_2$ topology of the bulk valence band.[10-13] The unique conducting surface states offer a new playground for studying the physics of quasiparticles with unusual dispersions, such as Dirac or Majorana fermions, as well as showing promising capabilities for spintronics (e.g. efficient spin-torque transfer). Experimentally, ever since the novel 3D TI family, including $Bi_2Se_3$, $Bi_2Te_3$, and $Sb_2Te_3$, was developed, the transport and optical properties related to topological surface states (SS) have been intensively studied.[11-16] In particular, a quantized anomalous Hall effect (QAHE) was found in a magnetic-ion-doped 3D TI thin film that presented a transverse current with extremely low dispersion.[17,18] To achieve the QAHE, both the linearly dispersed topological surface states and ferromagnetic ordering are needed. More recently, X-ray magnetic circular dichroism (XMCD) results[19] have demonstrated that the ferromagnetic ordering of spin-spin related surface states survives better than in the bulk states, which might imply that the critical temperature of the surface states is the key to QAHE.

Since the surface states of a TI are very important, it is a great challenge to develop a system in which surface-dominated transport survives at *high temperature*. In fact, most of the known TI materials are not bulk-insulating, which hinders study of the transport properties of the surface states. Therefore, bulk-insulating TIs with high resistivity are still required to extend the topological states to the high temperature region. A good example of a wide-gap TI system is $Bi_{2-x}Sb_xTe_{3-y}Se_y$ (BSTS)[10,11,20-23], which shows surface states that dominate the transport behavior at and below 30 K[10]. Since in bulk-insulating TIs, the Fermi surface is formed by pure Dirac dispersed surface states, the Shubnikov-de Haas (SdH) oscillations are strong evidence that can be used to evaluate the contribution of surface states. Tracing the SdH oscillations is an effective method to study the surface states in bulk-insulating 3D TIs. For example, the angular dependence of the SdH oscillations is key evidence that can be used to isolate surface-features. The oscillation magnitudes also provides insight into other properties of the electronic structure such as the effective mass. In realistic systems, the effective mass of a Dirac carrier is always nonzero instead of the ideal zero mass, and this causes temperature-induced damping of the amplitudes of SdH oscillations. One strategy that improves on BSTS, for which the surface band is not ideally linear, is to utilize a different TI: Sn-doped $Bi_{2-x}Sb_xTe_2S$, which has a wide bulk band gap [24-27]. In the latter compound, 2% tin is introduced to stabilize the crystal structure, and the corresponding material shows surface states with a linear dispersion over a large energy range.[24] We therefore focus on further optimization of this particular system by adding V at the bismuth sites to tune the bulk band gap. In this work, we found that the SdH oscillations could survive up to 50 K in the $V_{0.04}$ sample ($V_x$:$Bi_{1.08-x}Sn_{0.02}Sb_{0.9}Te_2S$, where $x = 0.04$), which allows 3D TI-based study of this substrate above the 50 K region. A special feature of this material is that the SdH oscillations are found in 1 mm-thick bulk crystals, which present useable surface states, making these crystals suitable for the fabrication of further van der Waals heterostructures.

**Results & Discussion**

A single crystal exfoliated from the as-grown $V_{0.04}$ ingot is shown in Fig. 1(a). The V:BSSTS single crystal shares the same crystal symmetry as its parent compounds $Sb_2Te_3$ and $Bi_2Te_3$, which belong to the *R-3m* space group, with quintuple layers piled up along the hexagonal *c*-axis (Fig. 1b). Traditionally, Se is widely used as a dopant to tune the band structure, which results in the (Bi,Sb)$_2$(Te,Se)$_3$ (BSTS) formula for the most popular topological insulator. Nevertheless, the defect chemistry limits our ability to grow large high-quality bulk single crystals of BSTS. Ideally, the new low-energy electronics industry requires wafer-size thick topological insulator materials with stable surface states and a large bulk band gap. Previous work by Kushwaha et al. demonstrated that Sn-doped $Bi_{1.1}Sb_{0.9}Te_2S$ single crystal has a low carrier concentration with clean surface states at low temperatures. To further optimize this material, we have employed V as a Bi-site dopant, which, we discovered, has the effect of making the bulk states more insulating (as sketched in Fig. 1(d)), whilst having minimal effects on the surface states. This is a major step forward towards realizing an ideal TI material for the electronics.

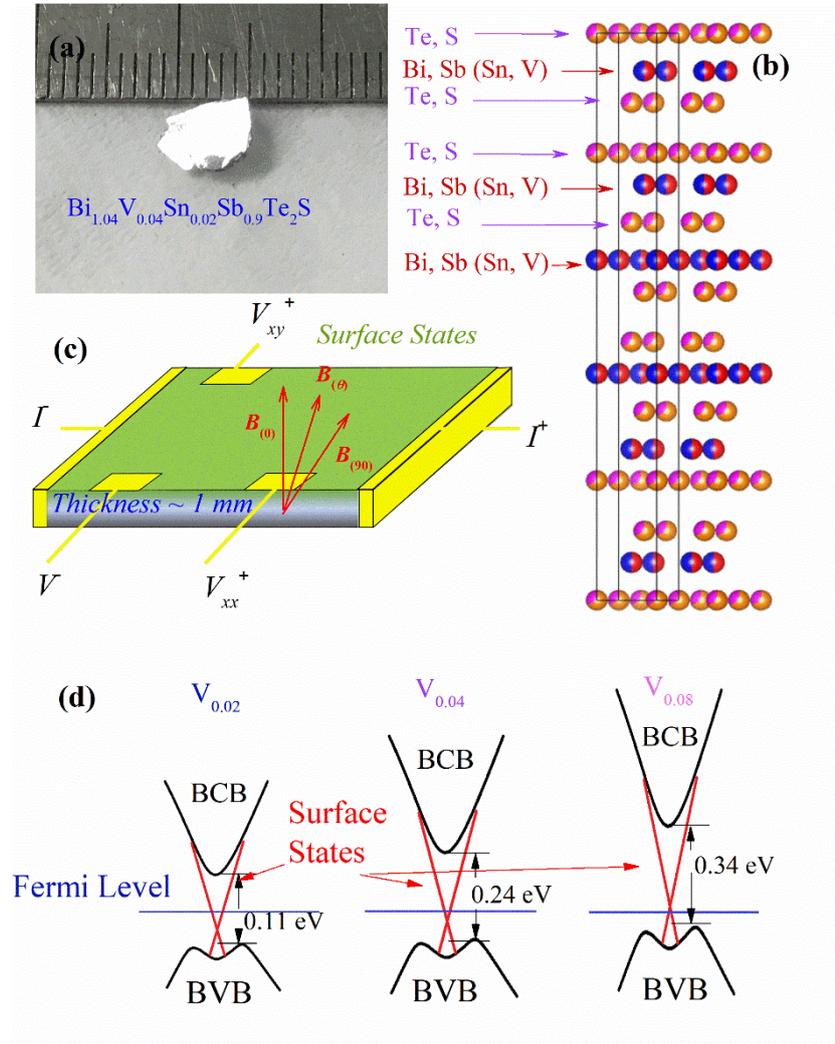

Fig. 1 Atomic and electronic structures of single crystal V:BSSTS. (a) Typical size of V$_{0.04}$ doped single crystal exfoliated from the as-grown ingot. (b) A sketch of the van der Waals layers in the V:BSSTS single crystal. (c) Schematic diagram of the transport measurement geometry for measuring a freshly-cleaved V:BSSTS surface. Note that, during tilting, the magnetic field is always perpendicular to the electronic current. (d) Possible bulk band structures in V:BSSTS single crystals.

Plots of the temperature dependence of bulk resistivity are presented in Fig. 2(a) for single crystal samples ~1 mm in thickness with different V doping levels. The resistivity curves for all of the vanadium doped samples show a steep upturn as the sample is cooled, indicating that the bulk V:BSSTS samples have become highly insulating. The residual resistance ratios (RRR), defined by RRR = $\rho(3\ K)/\rho(300\ K)$ for the samples, where $\rho$ is the resistivity, were 2.1, 9.4, and 127 for $V_{0.02}$, $V_{0.04}$, and $V_{0.08}$, respectively. As the V-doping level increases, the resistivity and the RRR increase dramatically, showing better insulating behavior, e.g., the maximum resistivity of the $V_{0.02}$ sample is near 0.1 $\Omega \cdot cm$, but it can reach about 0.7 and 10 $\Omega \cdot cm$ in the $V_{0.04}$ and $V_{0.08}$ samples, respectively. In the low temperature region below 10 K, the resistivity curves of the $V_{0.02}$ and -$V_{0.04}$ samples show an additional minor feature and a slight upturn, although this is negligible compared to the major transition at ~ 100 K. In the more highly insulating $V_{0.08}$ sample, the low temperature resistivity slightly drops with cooling. The temperature at which the surface states finally dominate the bulk resistivity depends on intrinsic factors, but also varies with V doping. The $\ln\rho$ versus $T^{-1}$ plots (shown in Fig. 1b) exhibit activated behavior, with the activation energy for transport at 110, 240 and 340 meV in the $V_{0.02}$, $V_{0.04}$, and $V_{0.08}$ samples, respectively.

We conducted measurements to determine the magnetoresistance (MR) ratio, MR = $(R(H)-R(0))/R(0)$, for all three V-doped samples in the low temperature region in order to understand the magnetotransport properties. In the $V_{0.02}$ sample, the MR curves first decrease in the low magnetic field region, and then increase with increasing magnetic field, up to about 16% at 3 K and 14 T. The maximum MR value at 14 T increases upon heating, so that it reaches 25% at 50 K. In the $V_{0.04}$ and $V_{0.08}$ samples, the MR values are always positive at all temperatures and under all magnetic field conditions, but they show temperature-damping behavior, with maximum MR of 50% and 110% at 3 K, respectively. Although the single crystals are similar in size and shape, the MR curves for the 3 samples are quite different from one another. In the $V_{0.02}$ crystal, the MR curves show a non-monotonic relationship with increasing magnetic field. Between 0 – 1 T, the 3 K MR curves slightly decrease to about -0.7%, and they then show a weak field dependence until about 5 T, after which, obvious oscillation patterns are displayed in the quasi-parabolic curves. The aforementioned oscillations, denoted as Shubnikov-de Haas (SdH) oscillations, result from Landau quantization. Upon cooling to different temperatures, the phenomenon of low-field MR decrease shows a similar tendency, but with a larger change in the relative drop, e.g., the minimum MR value (appearing at 1 T) at 5 K is about 1%, but it is near 1.6% at 8 – 50 K. Unlike the $V_{0.02}$ sample, the $V_{0.04}$ sample becomes more resistive with increasing field. Starting in the low-field region, the MR curves show a rapid increase at 3 K. Note that the total behavior of the $V_{0.04}$ sample is linear-like with high field SdH oscillations, but it shows a faster-than-linear increase in applied fields below 0.5 T fields. This faster rate of increase is attributed to the contribution of surface conduction, denoted as the weak anti-localization effect.[22] During heating, the low field behavior remains, but it becomes less significant and vanishes at 50 K. In the $V_{0.08}$ sample, the MR shows a simple linear increase with magnetic field in the $H < 6$ T region, after which, the SdH oscillations occur and the MR values increase more slowly than the linear tendency before.

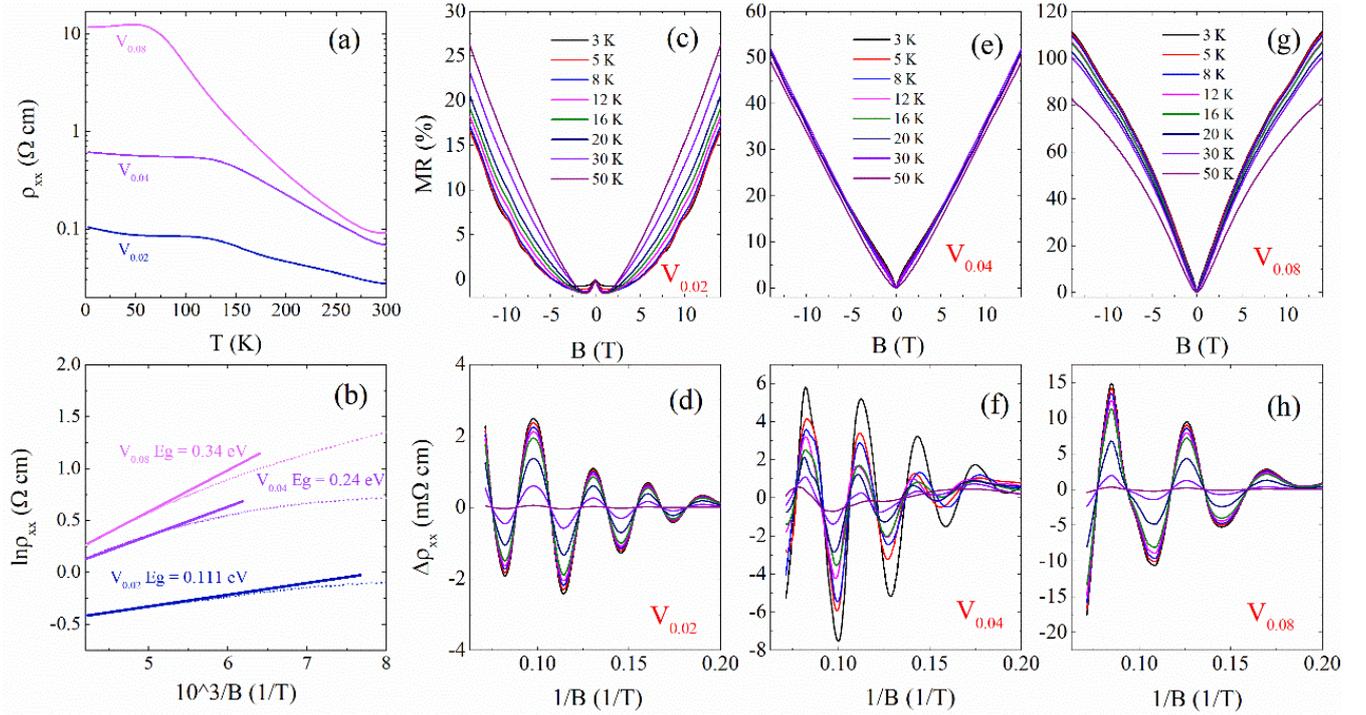

Fig. 2 Temperature dependence of the resistivity and the magnetoresistance (MR) of V:BSSTS. (a) The temperature dependence of the resistivity for crystals with different V contents. (b) Natural logarithm of the resistivity plotted against inverse temperature for different doping levels, linearly fitted to determine the band gaps. (c, e, g) The magnetoresistance at 3 – 50 K for crystals with different V doping levels. (d, f, h) SdH oscillation patterns obtained from the MR curves and plotted as a function of 1/$B$.

Let's then focus on the SdH oscillations in the MR curves, which are obtained via subtracting the smooth background and are plotted in Fig. 2(d, f, h). Since the bulk states are strongly insulating, the oscillation related Fermi surface is due to the contributions of the pure surface states. The oscillation patterns of each sample are almost the same, but with different amplitudes, which are damped during heating. Note that, in the $V_{0.04}$ sample, the SdH oscillations survive at 50 K, which means that the surfaces states are quite mobile at this temperature and might still exist at even higher temperatures.

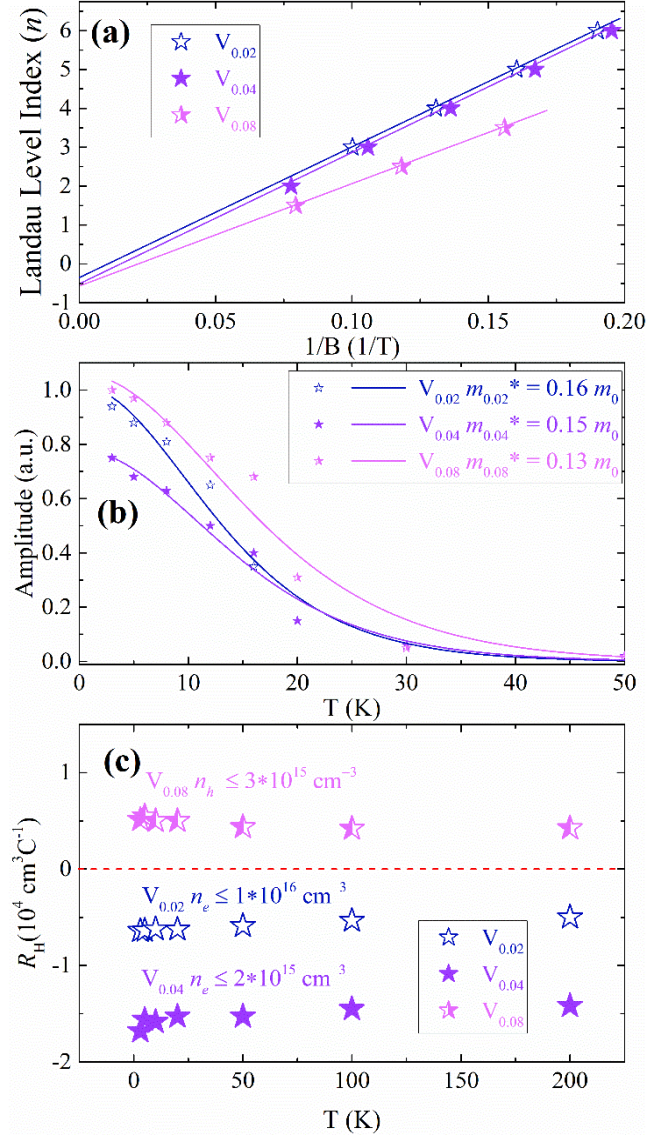

Fig. 3 Carrier analyses of the single crystals with different V doping levels. (a) Landau fan diagram related to the oscillation patterns at 3 K. (b) Fitting of the oscillation amplitudes by the LK formula. (c) Hall coefficients of single crystals with different V contents from 3 to 200 K.

The SdH oscillations for a semimetal can be described by the Lifshitz-Kosevich (LK) formula, with a Berry phase being taken into account for the topological system:

$$\frac{\Delta\rho}{\rho(0)} = \frac{5}{2}\left(\frac{B}{2F}\right)^{\frac{1}{2}} R_T R_D R_S \cos\left(2\pi\left(\frac{F}{B} + \gamma - \delta\right)\right)$$

Where $R_T = \alpha T \mu / B \sinh(\alpha T \mu / B)$, $R_D = \exp(-\alpha T_D \mu / B)$, and $R_S = \cos(\alpha g \mu / 2)$. Here, $\mu =$

$m^*/m_0$ is the ratio of the effective cyclotron mass $m^*$ to the free electron mass $m_0$; $g$ is the g-factor; $T_D$ is the Dingle temperature; and $\alpha = (2\pi^2 k_B m_0)/\hbar e$, where $k_B$ is Boltzmann's constant, $\hbar$ is the reduced Planck's constant, and $e$ is the elementary charge. The oscillation of $\Delta\rho$ is described by the cosine term with a phase factor $\gamma - \delta$, in which $\gamma = 1/2 - \Phi_B/2$, where $\Phi_B$ is the Berry phase. From the LK formula, the effective mass of carriers contributing to the SdH effect can be obtained through fitting the temperature dependence of the oscillation amplitude to the thermal damping factor $R_T$. From the temperature damping relationship, we obtain the Dingle temperatures of the three samples: 4.7, 6.5, and 9.4 K for $V_{0.02}$, $V_{0.04}$, and $V_{0.08}$, respectively. The effective masses for these crystals are 0.16 $m_0$, 0.15 $m_0$, and 0.13 $m_0$, respectively, as shown in Fig. 3(c). The quantum relaxation time and quantum mobility can also be obtained by $\tau = \hbar/2\pi k_B T_D$ and $\mu = e\tau/m^*$, respectively. According to the Onsager-Lifshitz equation, the frequency of quantum oscillation, $F = (\varphi_0/2\pi^2)A_F$, where $A_F$ is the extremal area of the cross-section of the Fermi surface perpendicular to the magnetic field, and $\varphi_0$ is the magnetic flux quantum. The cross-sections related to the 32, 30, and 25 T pockets are 0.34, 0.31, and 0.26 × 10$^{-3}$ Å$^{-2}$, respectively. The obtained parameters are summarized in Table 1 below.

The Berry phase can be obtained via the Landau fan diagram, which is shown in Fig. 3(b). Since $\rho_{xx} > \rho_{xy}$, we assign the maxima of the SdH oscillation to integer ($n$) Landau levels, to linearly fit the $n$ versus $1/H$ curve. As we can see in Fig. 3(b), the intercept has a value of around 0.5 for all three samples, which implies that these oscillations are contributed by the topological surface states.

| sample | $F$ (T) | $m^*$ ($m_e$) | $T_D$ (K) | $n$ ($10^{12}$ cm$^{-2}$) | $\mu$ ($10^4$ cm2·V$^{-1}$s$^{-1}$) |
|---|---|---|---|---|---|
| $V_{0.02}$ | 32 | 0.16 | 4.7 | 0.77 | 2.6 |
| $V_{0.04}$ | 30 | 0.15 | 6.5 | 0.69 | 1.2 |
| $V_{0.08}$ | 25 | 0.13 | 9.4 | 0.52 | 1.0 |

Table 1. Parameters determined from the SdH oscillations. $F$ is the frequency of quantum oscillation. $m^*$ is the effective mass. $T_D$ is the Dingle temperature. The 2D carrier's density and mobility are denoted by $n$ and $\mu$, respectively.

The SdH oscillation frequency is proportional to the area of the Fermi surface. We analyzed the oscillations of the three samples at 3 K and found that the SdH oscillation frequencies decreased with increasing vanadium concentration. We deduced that the vanadium doping changes the chemical potential of the samples, thus causing the Fermi level to shift downwards slightly towards the valence band, although remaining near the Dirac point in all cases. We further measured the Hall effect for all three samples, from which we calculated the Hall coefficient, as shown in Fig 3(c), where it is plotted versus temperature. In the low field region, the Hall resistivity curves increase/decrease linearly with the external magnetic field, where the Hall coefficient is positive or negative, respectively. The positive or negative value of the Hall coefficient depends on the carrier type in the samples. Interestingly, the carrier type and density in V:BSSTS changes with doping: in the $V_{0.02}$ and $V_{0.04}$ samples, it is electrons that contribute to the transport behavior, as in pure BSSTS; in contrast, the $V_{0.08}$ sample has *p*-type carriers. From the Hall coefficient, we can calculate that the carrier density is quite small in all three samples ($10^{15\text{-}16}$ cm$^{-3}$). Via V doping, we observe that the carrier type and density show large changes, which results in the different transport behavior.

To further understand the Fermi surface geometry, we performed angle-dependent SdH oscillation measurements. A good example of the angular dependence of the MR properties is shown in Fig. 4a, which is for the $V_{0.04}$ crystal. Note that, during rotation, the magnetic field is always perpendicular to the current, to make it possible to ignore the influence of the angle between the magnetic field and the current (as sketched in Fig. 1c). The rotation parameter $\theta$ is the angle between the *c*-axis and the direction of the magnetic field. At higher tilt angles, the MR value drops significantly, as along with the SdH oscillation amplitude. The oscillation patterns obtained from the MR curves at different orientations of the magnetic field are shown in Fig. 4(b-d) with the smooth background subtracted. One can see that the quantum oscillations become weaker in the rotation process and almost vanish at about 60 deg. We calculated each oscillation frequency, which are summarized in Fig. 4(e) and roughly fitted by the 1/cos$\theta$ relationship. The 2D-like relationship is strong evidence that the quantum oscillations are contributed by the topological surface states.

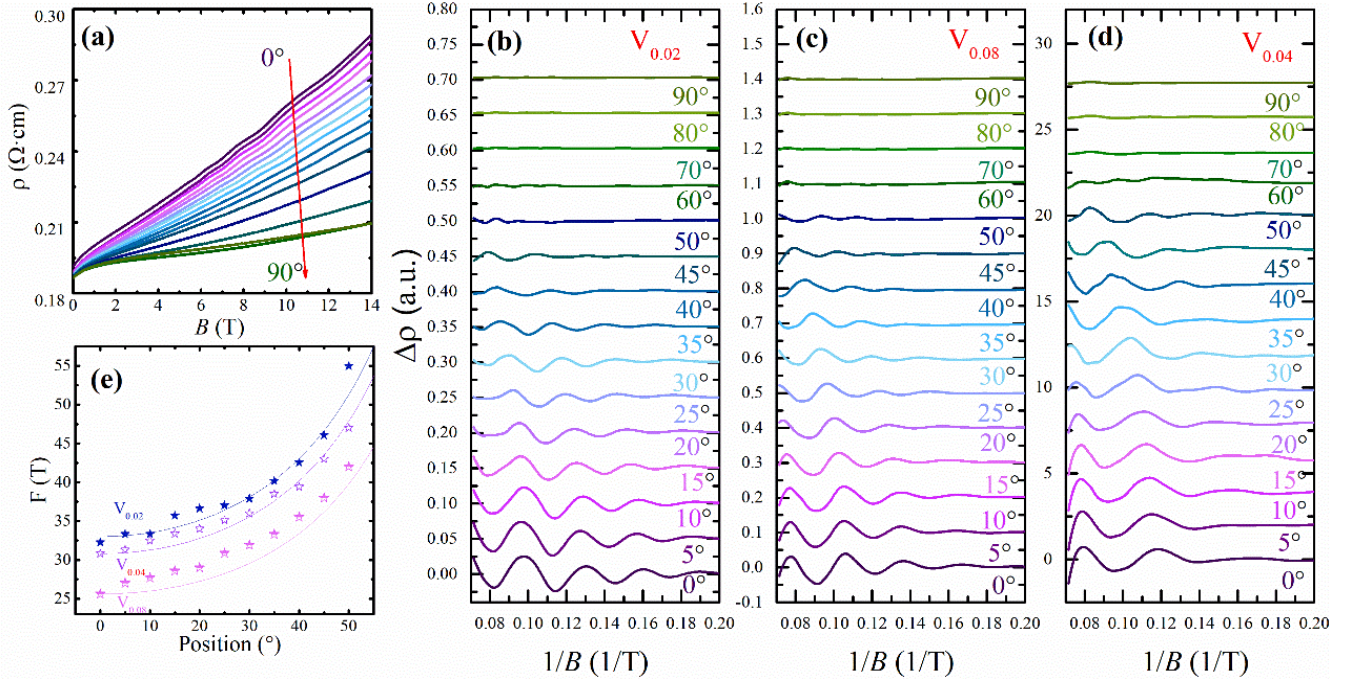

Fig. 4 Angle dependent SdH oscillations at 3 K. (a) Angular dependence of the MR for the $V_{0.04}$ sample at 3 K, where the applied field is tilted in the normal plane with respect to the current. (b-d) The oscillation patterns obtained from the angle dependent MR curves of the $V_{0.02}$, $V_{0.04}$, and $V_{0.08}$ single crystals, respectively. (e) Plots of the obtained oscillation frequencies vs. angle, which can be fitted into the two-dimensional $1/\cos(\theta)$ relationship.

**Conclusions**

We employed a simple-melting – slow-cooling method to grow single crystals of V, Sn doped $Bi_{1.1}Sb_{0.9}Te_2S$ single crystals. The *R-T* and Hall measurements show that all three samples are insulating in the 3-300 K region, with low carrier concentrations. SdH oscillations can be detected in the low temperature region, which implies that our insulating single crystals possess Fermi surfaces. Furthermore, we find strong 2D-like behavior and a $\pi$ Berry phase in all three samples, which provide compelling evidence that the bulks of these crystals are good insulators. Moreover, V dopant appears to tune both the type and the concentration of carriers in these TI systems, which provide us with an ideal platform to study their physical properties, as well as offering potential forMdevice fabrication related to their surface states.

**Methods**

To obtain bulk-insulating TIs, defect control is one of the most important factors in the crystal growth process. Here, we employ a simple melting-cooling method in a uniform-temperature vertical furnace to spontaneously crystallize the raw elements into a tetradymite structure ($V_x$:$Bi_{1.08-x}Sn_{0.02}Sb_{0.9}Te_2S$, V:BSSTS for short). Briefly, high-purity stoichiometric amounts (~10 g) of V, Bi, Sn, Sb, Te, and S powders were mixed via ball milling and sealed in a quartz tube as starting materials. The crystal growth was carried out using the following procedures: i) Heating the mixed powders to completely melt them; ii) Maintaining this temperature for 24 h to ensure that the melt is uniform; and iii) Slowly cooling down to 500 °C to crystallize the sample. After growth, single crystal flakes with a typical size of $5 \times 5 \times 1$ mm$^3$ could be easily exfoliated mechanically from the ingot. Naturally, the single crystals prefer to cleave along the [001] direction, resulting in the normal direction of these flakes being [001].

The electronic transport properties were measured using a physical properties measurement system (PPMS-14T, Quantum Design). Hall-bar contact measurements were performed on a freshly cleaved *ab* plane using silver paste cured at room temperature. The electric current was parallel to the *ab* plane while the magnetic field was perpendicular to the *ab* plane. The angle dependence of the magnetoresistance (MR) was also measured using a standard horizontal rotational rig mounted on the PPMS.